\documentstyle[twoside,fleqn,psfig,espcrc2]{article}%
\begin{document}
\renewcommand{\refname}{\normalsize\bf References}
\title{%
Structure of Quantum Chaotic Wavefunctions:
Ergodicity, Localization, and Transport}

\author{L. Kaplan%
\address{Institute for Nuclear Theory and Department of Physics,\\ 
University of Washington, Seattle, Washington 98195, USA}%
\thanks{lkaplan@phys.washington.edu}%
}
%
%
\begin{abstract}
\hrule
\mbox{}\\[-0.2cm]

\noindent{\bf Abstract}\\

We discuss recent developments in the study of quantum wavefunctions
and transport in classically ergodic systems. Surprisingly,
short-time classical dynamics leaves permanent imprints on long-time
and stationary quantum behavior, which are absent from the long-time
classical motion. These imprints can lead to quantum behavior on
single-wavelength or single-channel scales which are very different
from random matrix theory expectations. Robust and quantitative predictions
are obtained using semiclassical methods. Applications to wavefunction
intensity statistics and to resonances in open systems are discussed.\\[0.2cm]
\hrule
\end{abstract}

\maketitle

\section{Introduction}

The structure of quantum wavefunctions and the closely related problem
of quantum transport in classically non-integrable systems have received much
attention recently from a variety of physics communities. Questions concerning
the quantum behavior of systems with a generic classical limit are
of course of great fundamental interest; they are also very relevant not
only for nanostructure and mesoscopics experiments~\cite{meso},
but also for understanding
phenomena in areas as diverse as atomic physics~\cite{atom},
molecular and chemical
physics~\cite{chem}, microwave physics~\cite{microwave},
nuclear physics~\cite{nuclear}, and optics~\cite{optics}. Combined with
knowledge about spectral properties, wavefunction information can be used
to address conductance curves, susceptibilities, resonance statistics,
and delay times
in ballistic quantum dots. Similar wavefunction and
transport issues arise in other fields,
in the study of resonance statistics in microwave cavities,
photoionization cross sections, chemical
reaction rates, spectra of Rydberg atoms, lifetimes and emission intensities
for resonant optical cavities, and $S$-matrix properties in many systems.

In the classically integrable case, quantum wavefunctions are known to
be associated with the invariant tori of the corresponding classical
dynamics, satisfying the Einstein--Brillouin--Keller (EBK) quantization
conditions~\cite{ebk}.
Classical--quantum correspondence in the ergodic case is, however,
more subtle. Here, the typical classical trajectory uniformly visits
all of the energetically available phase space, so naively the typical
quantum wavefunction should also have uniform amplitude over an entire energy
hypersurface, up to the inevitable (Gaussian random) fluctuations. Such
behavior follows directly from Random Matrix Theory (RMT), which has been
proposed by Bohigas, Giannoni, and Schmit~\cite{bgs} to be the proper
description of quantum chaotic behavior in the semiclassical limit (i.e.
in the limit where the de Broglie wavelength $\lambda$
becomes small compared to the system size).
A similar conjecture by Berry~\cite{berry83} states that a typical
quantum wavefunction in this same limit should look locally like a random
superposition of plane waves of fixed energy, with momenta pointing in
all possible directions.

RMT~\cite{rmt},
a natural quantum analogue of classical ergodicity, turns out to 
describe well
{\it spectral properties on small energy scales} (e.g. the
distribution of nearest neighbor level spacings in the spectrum), but does
not always provide a valid description of {\it wavefunction structure
and transport} properties. Scars~\cite{citescars} provide one of the most
visually striking examples of strong deviation from RMT wavefunction
behavior; other examples include slow ergodic systems~\cite{wqe} and
Sinai-type systems~\cite{sinai}. In all these cases, non-RMT behavior can be
quantified, semiclassically predicted, and observed.

Section~\ref{erg} addresses the problem of wavefunction ergodicity
in broad terms. We are interested both in general questions concerning the
implications of global classical properties (such as ergodicity or mixing)
on quantum behavior and also in the way in which specific classical
structures (such as unstable periodic orbits) may leave their imprints on the
quantum eigenstates.  Several examples, including that of Sinai-type systems,
are discussed in Section~\ref{examp}.
Sections~\ref{scar1} and \ref{scar2} focus on the scar phenomenon and
quantitative predictions.
The overview format of this presentation requires
us to omit most derivations;
references to more detailed discussions in the literature can be found
throughout.

\section{Ergodic wavefunction structure and quantum transport}
\label{erg}

\subsection{Coarse-grained vs. microscopic ergodicity}

We must distinguish between two ways of extending classical notions
of ergodicity to the quantum case~\cite{wqe}.
First, we may consider wavefunction
intensity integrated over a classically defined region ${\cal R}$,
in the regime
where the wavelength $\lambda$
becomes small compared to the size of ${\cal R}$.
As can be shown rigorously~\cite{scdvz}, in this limit
the integrated intensity approaches a constant (equal to the area of
${\cal R}$ as a fraction of total phase space) for almost all wavefunctions. 
The result requires only long-time classical ergodicity and quantum--classical
correspondence at short times; a half-page physicists' derivation can be found
in ~\cite{quamb}. We note that this macroscopic or ``coarse--grained'' 
type of quantum ergodicity
is clearly implied by RMT, but is in fact a much weaker condition. Quantum
wavefunction structure on coarse--grained scales is relevant
for studying conductance and conductance fluctuations through wide,
multichannel leads, for fast decay processes, and generally for analyzing open
systems in the overlapping resonance regime.

RMT, on the other hand, is a prediction about wavefunction uniformity at
the quantum scale, i.e. on
the scale of a single wavelength (or single momentum channel, or most
generally at the scale of a single $\hbar$-sized cell in phase space). 
This kind of uniformity is a much stronger condition than coarse-grained
ergodicity, and several examples will be given later in this section
where microscopic quantum ergodicity is violated in classically ergodic
systems. Wavefunction structure at the quantum scale is relevant, obviously,
for transport through narrow (or tunneling) leads.

\subsection{Measures of microscopic ergodicity}

First we must define quantitative measures of ergodicity or localization at the
microscopic scale~\cite{wqe}. Let $|n\rangle$ be an $N-$dimensional
basis of eigenstates, and
$|a\rangle$ some localized test basis, e.g. position, momentum, or phase
space Gaussians, as is physically appropriate in a given system. For example,
in discussing Anderson-type lattice localization, we may choose our test basis
$|a\rangle$
to be the position basis, whereas in scattering problems plane waves may be
a more natural choice. Sometimes we take $|a\rangle$ to be the eigenstates
of a zeroth-order Hamiltonian $H_0$ or of a zeroth-order scattering
matrix $S_0$. We are then interested in the wavefunction intensities
$P_{an}=|\langle a|n\rangle|^2$
and their correlations. For simplicity of presentation we assume there
are no conserved quantities, so classically nothing prevents each eigenstate
$|n\rangle$ for overlapping equally with each $|a\rangle$ (of course,
the formalism
generalizes naturally to the more general case of classical symmetries,
see~\cite{classym}). We adopt the normalization convention
\begin{equation}
\langle P_{an} \rangle =1\,,
\end{equation}
where the averaging is done over test states $|a\rangle$, over
wavefunctions $|n\rangle$, or over an appropriate ensemble of systems.

The first nontrivial moment of the $P_{an}$ distribution is the
{\it inverse participation ratio}:
\begin{equation}
\label{ipra}
{\rm IPR}_a = {1 \over N} \sum_{n=1}^{N} P_{an}^2 \ge 1\,,
\end{equation}
which measures the mean squared wavefunction intensity at $|a\rangle$,
or, alternatively, the inverse fraction of wavefunctions having significant
intensity at $|a\rangle$. ${\rm IPR}_n$, the mean squared intensity of
a single wavefunction $|n\rangle$ averaged over position $|a\rangle$,
is defined analogously: it measures the inverse fraction of phase space
covered by $|n\rangle$. A global IPR measure may also be conveniently defined:
\begin{equation}
{\rm IPR}=  {1 \over N}\sum_{a=1}^{N} {\rm IPR}_a =
 {1 \over N}\sum_{n=1}^{N} {\rm IPR}_n \ge 1\,,
\end{equation}
and provides a simple one-number measure of the degree of localization
in a quantum system. An IPR of unity corresponds to perfect ergodicity
(each wavefunction having equal overlaps with all test states),
while ${\rm IPR}=N$ indicates the greatest possible degree of localization
(the wavefunctions $|n\rangle$ being identical with the test states
$|a\rangle$).

Besides being the first nontrivial moment of the intensity distribution, the IPR
measure is useful because of its connection with dynamics. For example,
${\rm IPR}_a$ is proportional to the averaged long-time return probability
for a particle launched initially in state $|a\rangle$:
\begin{equation}
\label{timedomipr}
{\rm IPR}_a= N\lim_{T \to \infty} {1 \over T} \sum_{t=0}^{T-1}
|\langle a |e^{-i \hat H t/\hbar} |a \rangle|^2 \,.
\end{equation}
It is intuitively clear that enhanced long-time return probability is
associated with increased localization. We may generalize ${\rm IPR}_a$ to a
transport or wavefunction correlation measure between local states
$|a\rangle$ and $|b\rangle$:
\begin{equation}
P_{ab}= \sum_{n=1}^{N}
{P_{an}P_{bn} \over N} =
\lim_{T \to \infty} {N \over T}\sum_{t=0}^{T-1}
|\langle a |e^{-i \hat H t/\hbar} |b \rangle|^2 \,.
\end{equation}
Of course, the mean ${1 \over N} \sum_bP_{ab}=1$ by unitarity; the
simplest nontrivial measure of transport efficiency is thus
\begin{equation}
\label{qa}
Q_a={1 \over N} \sum_b P_{ab}^2 \ge 1 \,,
\end{equation}
which measures the inverse of the fraction of phase space accessible from
$|a\rangle$ at long times. $Q_a=1$ for all $|a\rangle$
(the RMT result for $N \to \infty$) indicates perfect long-time transport,
and vanishing wavefunction correlations.

\subsection{Examples}
\label{examp}

The Random Matrix Theory description
is free of all dynamical information about the system under study and
thus can hardly be expected to provide a correct statistical
description of all quantum behavior. It serves, however, as a very useful
baseline with which real quantum chaotic behavior may be compared.
In RMT, the wavefunction intensities $P_{an}$ are squares of Gaussian
random variables and so are drawn from a $\chi^2$
distribution (of one degree of freedom for real
overlaps $\langle a|n\rangle$ or two for complex overlaps),
with a mean value of unity.
This easily leads to ${\rm IPR}_{\rm RMT}=3$ in the real case or $2$ in the
complex case. Notice that even in RMT, quantum fluctuations cause
wavefunctions to be less ergodic than the classical expectation
${\rm IPR}_{\rm Clas}=1$. By quantum localization, however, we always
mean fluctuation
in the intensities {\it in excess
of what would be expected from a Gaussian random
model,} i.e. ${\rm IPR} > {\rm IPR}_{\rm RMT}$. Examples abound of such
anomalous quantum behavior in classically ergodic systems, and several
are described below.
We also mention here that in RMT, the channel--to--channel
transport efficiency $P_{ab}$ approaches
unity for all channels $|a\rangle$, $|b\rangle$ in the $N\to \infty$
semiclassical limit.

(i) {\it Scarring} is the anomalous enhancement or suppression of quantum
wavefunction intensity on the {\it unstable}
periodic orbits of the corresponding classical
system. This localization behavior is perhaps surprising from a naive
classical point of
view, since in the time domain (see Eq.~\ref{timedomipr}) it implies an
enhanced long-time return probability for a wavepacket launched on an
unstable periodic orbit. Paradoxically, this is in contrast with the classical
behavior, where a probability distribution spreads itself evenly over the
entire ergodic space at long times and retains no memory of its initial
state. Quantum long-time dynamics, which of course contains phase
information,
thus retains a much {\it better} memory of the short-time classical behavior
than does the long-time classical dynamics, for arbitrarily small values of
$\hbar$.
Specifically, for a wavepacket $|a\rangle$ optimally oriented on
an orbit of instability exponent $\beta$, we have~\cite{citescars}
\begin{eqnarray}
{\rm IPR}_{\rm Scar} &=& \left[\sum_{m=-\infty}^{\infty}
 { 1\over \cosh{\beta m}}\right] {\rm IPR}_{\rm RMT} \\ & \to & 
{\pi \over \beta}  {\rm IPR}_{\rm RMT} \gg {\rm IPR}_{\rm RMT} \,,
\end{eqnarray}
where the limiting form is valid for weakly unstable orbits ($\beta \ll 1$).
A finite fraction of order $\beta$ of all wavefunctions are {\it scarred}:
they have intensity $O(\beta^{-1})$ on the periodic orbit compared with
the mean intensity, while most of the remaining wavefunctions
are {\it antiscarred}, and have intensities on the orbit
as small as $\exp{(-\pi^2/2\beta)}$ compared with the mean.
The source of this localization effect will be outlined in Section~\ref{scar1}.
The transport measure $P_{ab}$ is also affected by scarring: if
both leads $|a\rangle$ and $|b\rangle$ are located near periodic orbits, 
$P_{ab}$ can be enhanced or suppressed depending on whether the two
orbits are ``in-phase" or ``out-of-phase" in the range of energies considered.
These long-time transport results are $\hbar-$independent and are obtained
directly from the short-time linearized classical dynamics.

(ii) {\it The tilted wall billiard}
(a rectangular box with one wall tilted relative
to the other three), is a classically ergodic system in which the wavefunctions
(as measured by the IPR) become {\it less and less ergodic} in the classical
$\hbar \to 0$ (or $\lambda \to 0$)
limit~\cite{wqe}. (This kind of behavior is possible 
because the $\hbar \to 0$ limit fails to commute with the
infinite time $t \to \infty$ limit.) Specifically, for wavefunctions expressed
in channel (momentum) space, we have
\begin{equation}
{\rm IPR}_{\rm Tilted}  \sim \lambda^{-1/2} / \log \lambda^{-1} \to \infty\,.
\end{equation}
This anomalous wavefunction behavior at the single-channel scale is entirely
consistent with ergodicity on coarse-grained scales; coarse-graining
over $\sim \lambda^{-1} / \log \lambda^{-1}$ channels is required to retrieve
the classical behavior. Quantum transport in these systems is also
anomalous and is dominated by diffractive effects (due to the slowness
of classical phase space exploration).

(iii) As a final example we mention {\it the Sinai billiard,}
a paradigm of classical
chaos~\cite{classinai}. This consists of a circular
obstruction of
diameter $d \gg \lambda$ placed inside a rectangle. Again using momentum
channels for the test basis $|a\rangle$, we obtain~\cite{sinai}
\begin{equation}
{\rm IPR}_{\rm Sinai}  \sim (\log \lambda^{-1}) / d \to \infty
\end{equation}
for fixed $d$ in the $\lambda \to 0$ limit. Again, we see nonergodic
$\lambda \to 0$ wavefunctions in a classically ergodic system. The
distributions of
wavefunction intensities $P_{an}$ and IPR's (${\rm IPR}_a$ and
${\rm IPR}_n$, see Eq.~\ref{ipra})  all display power-law
tails (in contrast with the exponential tail prediction of RMT).
Transport is similarly anomalous: for the typical channel 
$|a\rangle$, $Q_a \sim d^{-1}$ (see Eq.~\ref{qa}); i.e. at long times
a given channel is coupled to only a fraction $\sim d \ll 1$ of all other
channels.

\section{Scars of quantum chaotic wavefunctions}
\label{scar1}

\subsection{Phenomenology and conceptual issues}

Scars have been observed experimentally and numerically in a wide variety of
systems. These include semiconductor heterostructures, where scars
are observed to affect tunneling rates from a 2D electron gas 
into a quantum
well~\cite{semiscar}, microwave cavities~\cite{microscar},
the hydrogen atom in a uniform magnetic field~\cite{hydroscar},
acoustic radiation from membranes~\cite{membscar}, and the stadium billiard,
for which wavefunctions as high as the millionth state in the spectrum
can be numerically analyzed~\cite{stadscar}. Nevertheless, confusion over
the definition and measures of scars (and the dearth of quantitative
predictions) have until recently led some to question the existence of a scar
theory, and even to doubt the survival of the phenomenon in the classical
limit. Recent theoretical developments~\cite{citescars} enable us to make
robust, quantitative predictions about how strongly a given orbit will be
scarred and how often, as a function of energy and other system parameters;
real comparison with numerical and experimental data is therefore made
possible. We also note that statistical and quantitative data, rather than
anecdotal (in the form of wavefunction plots) is necessary as a test of scar
theory or of any other theory of a localization phenomenon, since considerable
fluctuations of wavefunction intensity occur even in the context of RMT
(see discussion in first paragraph of Section~\ref{examp}).

We address some common misconceptions and summarize key facts about
the scar phenomenon: (i) Scarring is associated with
{\it unstable} periodic orbits, not with stable or marginally stable ones,
though these of course also do attract wavefunction intensity; a qualitative
difference arises because of classical--quantum noncorrespondence in the
unstable case (Section~\ref{examp}(i)). (ii) Weak scars are not always visible
to the naked eye; scars are defined and measured according to a statistical
definition. (iii) Scarring predictions are robust, valid even when the exact
dynamics is not known well enough to allow individual eigenstates to be
determined either quantum mechanically or semiclassically.
(iv) The amount of scarring associated with individual eigenfunctions
varies significantly from state to state, but in accordance with a
theoretically predicted distribution. (v) If an optimally chosen phase space
basis is used, the typical intensity enhancement factor as well
as the full distribution of scar intensities  can be given as a function of the
instability exponent $\beta$.

\subsection{Short-time effects on stationary properties}

We define the autocorrelation function for test state $|a\rangle$:
\begin{equation}
A(t)=\langle a|\exp[-iHt]|a\rangle\,,
\end{equation}
the Fourier transform of which is the local density of states (LDOS)
\begin{equation}
\label{spec}
S(E)=\sum_n P_{an} \delta(E-E_n)\,.
\end{equation}
Then, if we know the statistical properties of the return amplitude $A(t)$,
we also know the statistical properties of the LDOS at  $|a\rangle$, and
specifically of the wavefunction intensities $P_{an}$
(see e.g. Eq.~\ref{timedomipr}). Furthermore, given information about the
return amplitude $A(t)$ for short times only (say for $|t|<T_0$), we
immediately obtain the LDOS envelope $S_{\rm smooth}(E)$, which is the 
energy-smoothed (on scale $\hbar/T_0$) version of the true line-spectrum
$S(E)$. So large short-time recurrences in $A(t)$ get `burned into' the
spectrum. 

Specifically, if $|a\rangle$ is a minimum uncertainty wavepacket
optimally placed on a periodic orbit of period $P$, then
at integer multiples $t=mP$ of this period, by Gaussian integration we
easily obtain~\cite{citescars}
\begin{equation}
A_{\rm short}(t)=\exp[imS/\hbar]/\sqrt{\cosh \beta m}\,,
\end{equation}
where the classical orbit has action $S$ and instability exponent $\beta$.
For small $\beta$, the Fourier transformed spectral envelope
$S_{\rm smooth}(E)$ has bumps of width scaling as $\beta/P$ and height 
scaling as $\beta^{-1}$. Now after the mixing time ($\sim \beta^{-1}
\log N$), escaping probability starts returning to the origin, leading
to fluctuations in the spectral envelopes, and eventually to
discrete delta-function peaks (Eq.~\ref{spec}). The nonlinear scar theory
allows the long-time returning amplitude to be analyzed as a homoclinic
orbit sum~\cite{citescars}, and leads to the prediction that the
spectral intensities are given by $\chi^2$ variables multiplying the short-time 
envelope:
\begin{equation}
P_{an} = |r_{an}|^2 S_{\rm smooth}(E_n)\,,
\end{equation}
where the $r_{an}$ follow a Gaussian random distribution of variance unity.

\section{Quantitative predictions for scar effects}
\label{scar2}

\subsection{Wavefunction intensity statistics}
\begin{figure}
\centerline{
\psfig{file=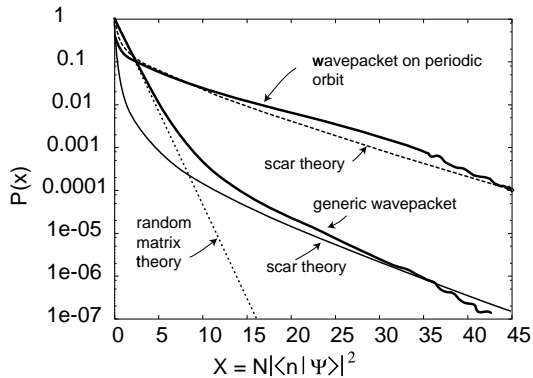,width=3.2in}}
\caption{Cumulative wavefunction intensity distribution (a) as
measured by a Gaussian test state optimally centered on a periodic
orbit with instability exponent $\beta=\log 2$, plotted as the upper thick
curve with scar theory prediction for the tail
(Eq.~\ref{th1}) given by dashed curve,
and (b) averaged over the entire phase space of size $200h$, plotted
as lower thick curve with theory for the tail
(Eq.~\ref{th2}) given by solid curve.
The dotted line is the RMT exponential prediction
(after~\protect\cite{wavefcn}).}
\label{figscar}
\end{figure}

As a simple application, we may compute the distribution of wavefunction
intensities on a periodic orbit of instability exponent $\beta$~\cite{wavefcn}.
For complex
wavefunctions, the probability to have an intensity exceeding the mean
by a factor $x$ is given by $\exp(-x)$ in RMT; on a periodic orbit the
tail of the distribution is instead given by
\begin{equation}
P(\beta,x)=C \beta (\beta x)^{-1/2} e^{-\beta x/Q}\,,
\label{th1}
\end{equation}
where $C$ and $Q$ are known numerical constants. Similarly, the tail
of the overall intensity distribution (sampled over the entire phase
space) is given by
\begin{equation}
P(\beta,x)=C' \beta\hbar (\beta x)^{-3/2} e^{-\beta x/Q}\,,
\label{th2}
\end{equation}
where $\beta$ is now the exponent of the least unstable periodic orbit.
These two results are illustrated in Fig.~\ref{figscar}.
Finally, for an ensemble of
systems, a power-law tail is obtained, in contrast with the exponential 
prediction of RMT, and large intensities have been observed
numerically with frequency exceeding by $\sim 10^{30}$ the RMT predictions.

\subsection{Enhancement in probability to remain}
\label{subenh}

\begin{figure}
\centerline{
\psfig{file=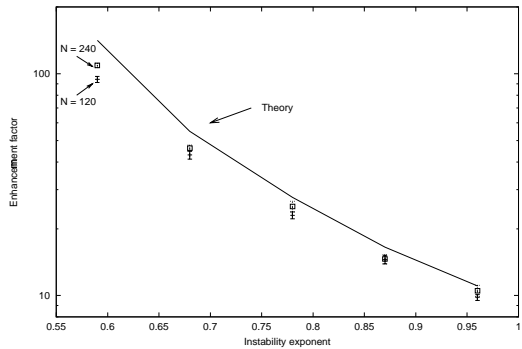,angle=270,width=3in}}
\caption{The long-time enhancement of the probability to remain in
a weakly open system when the tunneling lead is placed on a periodic orbit
is plotted as a function of the instability exponent $\beta$ of the orbit.
Data is shown for system sizes $N=120$ (plusses) and $N=240$
(squares). The $N \to \infty$ theoretical prediction
(Section~\ref{subenh}), growing exponentially with decreasing 
$\beta$, is shown as a solid curve. For large $\beta$, the
enhancement factor converges to $1$, the RMT prediction
(after~\protect\cite{open}).}
\label{figenh}
\end{figure}

The above methods can also be applied to the open systems case~\cite{open}.
Consider
the probability to remain inside a chaotic quantum well
coupled to the outside via a tunneling lead. For a lead optimally
placed with respect to a periodic orbit of exponent $\beta \ll 1$, 
the long-time probability to remain in enhanced by a factor
$\exp(\pi^2/2 \beta)$ (see Fig.~\ref{figenh}),
compared with the RMT expectation. (The enhancement is due the
antiscarred states (Section~\ref{examp}(i)), which have very small
coupling to the lead.) Of course, the
classical probability to remain is exponential and independent of lead
position for a chaotic system with a narrow lead. So once again we see
long-time quantum mechanics retaining an imprint of short-time classical
structures which is absent from long-time classical behavior. The analysis
can of course be extended to the study of conductance peak statistics
in two lead systems, where one or both leads are located near short periodic
orbits (compare with the discussion of $P_{ab}$ statistics in
Section~\ref{examp}(i)).

\section{Conclusions}

We have seen that wavefunction structure and transport in classically
ergodic systems, including the paradigmatic Sinai billiard system,
can differ greatly from RMT expectations, and in fact
can deviate further and further from RMT in the semiclassical limit. This
non-ergodic quantum behavior at the scale of single wavelengths or single
quantum channels can be quantitatively and robustly predicted using only
short-time classical information. This anomalous small-scale quantum
behavior is also entirely consistent with ergodicy on coarse-grained
scales, as studied by Schnirelman, Zelditch, and Colin de Verdiere.

Scarring is a fascinating example of the influence of identifiable classical
structures on stationary quantum properties (e.g. eigenstates) and long-time
quantum transport in classically chaotic systems. Short unstable periodic
orbits leave a strong imprint on the long-time properties of the quantum
chaotic system, even though classical dynamics loses all memory of these
structures at long times. Scar theory makes robust and quantitatively
verified predictions about properties such as the wavefunction intensity
distribution in a chaotic system, including the power-law tail observed
after ensemble averaging. A lead centered on an unstable periodic orbit has
been shown to produce many exponentially narrow quantum resonances, even
though the decay times are very long compared with all other time scales in
the problem. Enhancement factors of $10$ to $100$ in the total probability
to remain are easily observed for
moderate values ($1.0$ to $0.5$) of the instability exponent.

\section{Acknowledgments}
It is a pleasure to thank E. J. Heller for many fruitful discussions.
This work was supported by the National Science Foundation under
Grant No. 66-701-7557-2-30.

\end{document}